\documentstyle[twocolumn,aps]{revtex}
\begin{document}
\draft
\title{\bf Delocalization of 2D Dirac Fermions: the Role of a Broken 
Supersymmetry}
%\author{[Preliminary Version]}
\author{K. Ziegler}
\address{Max-Planck-Institut f\"ur Physik Komplexer Systeme,
Au\ss enstelle Stuttgart, Postfach 800665, D-70506 Stuttgart, Germany}
%\address{and}
%\address{Institut f\"ur Physik, Universit\"at Augsburg,
%D-86135 Augsburg, Germany}
\date{\today}
\maketitle
\begin{abstract}
The mechanism of delocalization of two-dimensional Dirac fermions with random
mass is investigated, using a superfield representation. Although
localization effects are very strong, one fermion component can delocalize
due to the spontaneous breaking of a special supersymmetry of the model.
%The other fermions and the bosons become localized.
The delocalized fermion has a non-singular density of states and
is decribed by a diffusion propagator.
Supersymmetry is restored if the mean of the random mass is sufficiently
large. This is accompanied by a critical boson component.
\end{abstract}
\pacs{PACS numbers: 73.40.Hm, 71.55.Jv, 72.15.Rn, 73.20.Jc}
%%%%%%%%%%%%%%%%%%%% Introduction %%%%%%%%%%%%%%%%%%%%%%%%
The transition from localized to delocalized states of non-interacting quantum
particles in a random potential is a phenomenon which is characterized by
symmetries. In contrast to classical
critical phenomena, where symmetries are either discrete or {\it compact}
continuous, it was observed that the transition from localized to delocalized
states of a particle, described by a discrete Schr\"odinger equation
(tight-binding model), are related to {\it non-compact} symmetry groups
\cite{wegner}.
Nonlinear $\sigma$ models with the corresponding symmetries
provide an effective large scale description, presenting the relevant
degrees of freedom for localization or delocalization.
They describe an effective diffusion of the quantum
particle with diffusion coefficient $D\ge0$. An important physical property 
of $D$ in the nonlinear $\sigma$ model is its flow under renormalization. 
In general there are fixed points, one for delocalized states ($D>0$) and
one for localized states ($D=0$) \cite{abrahams}.
In two-dimensional systems the renormalization always
drives the diffusion coefficient to zero \cite{abrahams},
therefore reflecting the absence of delocalized states, at least
in the absence of more complicated extensions of the Schr\"odinger equation
like spin-orbit coupling.

It turned out that for a number of interesting physical systems the
effective quantum theory is not defined by Schr\"odinger particles but by
Dirac fermions. The main reason for this is a linear dispersion and a 
sub-structure, either given by a sublattice or a spin. For instance,
the integer quantum Hall transition (QHT) in a 2D electron gas with
magnetic field can be formulated with Dirac fermions without a magnetic field 
\cite{fi,haldane,ludwig,hats,osh,zie2,zie6,ho/chalker}. Other
examples for Dirac fermions are the degenerate semiconductor \cite{fradkin}
and quasiparticles in a 2D d-wave superconductor \cite{nersesyan,zie7}.

A Dirac fermion is a quantum particle with symmetry properties different
from those of the Schr\"odinger particles. In particular,
the symmetry of the 2D Dirac Hamiltonian is discrete in contrast to the
continuous symmetries of the Schr\"odinger Hamiltonian. This fact has
important consequences for the delocalization of the Dirac particle in
$d=2$ \cite{zie6}, and will be discussed in this letter.

%%%%%%%%%%%%%%%%%%%% The Model %%%%%%%%%%%%%%%%%%%%%%%%%%%
The Dirac Hamiltonian in 2D reads
\begin{equation}
H_D=i\nabla_1\sigma_1+i\nabla_2\sigma_2+M\sigma_3.
\label{hamiltonian}
\end{equation}
$\nabla_j$ is the lattice difference operator in $j$--direction, $M$ is the
mass of the particle and $\sigma_j$ is a Pauli matrix.
The localization properties of massless Dirac fermions with random vector
potential was recently studied \cite{ludwig,morita}. It turned out that the
low energy states are delocalized. Since the related Hamiltonian matrix has
only off-diagonal elements this result can be compared with similar
observations in 1D systems: there are delocalized states at the band-center 
with a singular density of states (DOS) if
the Hamiltonian represents hopping between sublattices or different spin
states \cite{lifshits}. In contrast to this it of interest to consider
models where also a diagonal (potential) term appears in the Hamiltonian,
and which have a non-singular DOS.

The Dirac Hamiltonian
$H_D$ is an effective two-particle Hamiltonian because the Dirac theory
includes particles and holes as the two components of the Dirac spinor.
$H_D$ is Hermitean and
invariant under the transformation $H_D\to-\sigma_3 H_D\sigma_3$, provided
the Dirac mass $M$ is zero. However, this symmetry is not interesting here
because it is always broken by the 
%random
mass. Moreover, there is a space-dependent discrete transformation
\begin{equation}
H_D\to -SH_D'S
\label{discrete}
\end{equation}
for which the massive $H_D$ is invariant.
The $2\times2$ matrix $S_r$
%=\sigma_1$ or $=\sigma_2$ 
is changing between
$\sigma_1$ and $\sigma_2$ by going from one site to its nearest neighbor site,
and $H_D'$ is obtained from $H_D$ by a space-rotation of $\pi/2$ and a
reflexion of the $y$-axis. (This is just an exchange of $\nabla_1$ and 
$\nabla_2$ in $H_D$.)

In order to compare the Dirac Hamiltonian with the
corresponding Hamiltonian $H=\nabla^2+V$ of a Schr\"odinger particle in a
random potential $V$, we extend the latter to $H_S=(\nabla^2+V)\sigma_3$.
This Hamiltonian describes particles {\it and} the corresponding holes, and
can be used to express the two-particle Green's function for Anderson
localization without a magnetic field. $H_S$ is symmetric and invariant
under a non-compact continuous 
%orthogonal
symmetry under $H_S\to (c\sigma_0+s_1\sigma_1+s_2\sigma_2)H_S
(c\sigma_0+s_1\sigma_1+s_2\sigma_2)$ with the condition $c^2-s_1^2-s_2^2=1$.
The role of the chemical potential in the case of Dirac particles is played by
the Dirac mass, as it was earlier discussed by Ludwig et al. \cite{ludwig}.

Transport properties can be evaluated from the two-particle Green's
function \cite{zie6}
\begin{equation}
K(r,r';\epsilon)\equiv
-\langle Tr_2[G(r,r';i\epsilon)\sigma_1G^T(r',r;i\epsilon)\sigma_1]\rangle,
\end{equation}
where $G(r,r';i\epsilon)\equiv (H+i\epsilon)^{-1}_{rr'}$ is the one-particle
Green's function of $H_D$ or $H_S$, and $\langle ...\rangle$ the 
average over random contributions in the Hamiltonian. For localized states 
the two-particle Green's function decays exponentially on the localization
length.

$H_S$ is invariant under the
transposition $^T$ of the matrix elements whereas the Dirac Hamiltonian
$H_D$ is not. It is convenient to write the
two-particle Green's function as a functional integral
\begin{eqnarray}
G_{jj'}(r,r';i\epsilon)G^T_{k'k}(r',r;i\epsilon) & = &
%\equiv(i\epsilon+H)^{-1}_{rj,r'j'}(i\epsilon+H^T)^{-1}_{r'k',rk}
\nonumber\\
\int\chi_{r'j'}{\bar \chi}_{rj} & \Psi_{rk} & {\bar \Psi}_{r'k'}\exp(-S_0)
%{\cal D}\Psi {\cal D}{\bar \Psi} {\cal D}\chi {\cal D}{\bar \chi}.
{\cal D}\Psi {\cal D}\chi .
\label{fint}
\end{eqnarray}
$S_0$ is a quadratic form of the four-component superfield $(\chi_r,\Psi_r)$
\begin{equation}
%S_0=
-i{\rm sign}(\epsilon)\sum_{r,r'}\pmatrix{
\chi_r\cr
\Psi_r\cr
}\cdot\pmatrix{
H+i\epsilon&0\cr
0&H^T+i\epsilon\cr
}_{r,r'}\pmatrix{
{\bar \chi}_{r'}\cr
{\bar \Psi}_{r'}\cr
}
\label{ssa0}
\end{equation}
with a complex component $\chi_r$ and a Grassmann component $\Psi_r$. The
reason for introducing the superfield is that an extra normalization
factor for the integral in Eq.(\ref{fint}) is avoided because of 
$\int\exp(-S_0){\cal D}\Psi {\cal D}\chi
=\det(H_D^T+i\epsilon)/\det(H_D+i\epsilon)=1$.
It is crucial that $S_0$
is {\it not} of the usual supersymmetric form \cite{efetov}, where both 
diagonal elements are $H+i\epsilon$, if $H^T\ne H$ \cite{note}.
This reflects a fundamental difference between the symmetric Schr\"odinger
Hamiltonian $H_S$ and the asymmetric Dirac Hamiltonian $H_D$ for the 
construction of collective fields. In the following we will concentrate on 
the Dirac Hamiltonian and refer to the literature for the case of the
Schr\"odinger Hamiltonian \cite{wegner,efetov,altshuler}.

In addition to the discrete symmetry of $H_D$ there is an invariance 
of the action $S_0$ for $\epsilon=0$ under supersymmetry transformation 
\begin{eqnarray}
{\bf H}_D\equiv\pmatrix{
H_D&0\cr
0&H_D^T\cr
}\to U\pmatrix{
H_D&0\cr
0&H_D^T\cr
}U={\bf H}_D
\nonumber\\
{\ \ \ \ }{\rm for}\ \ U = \pmatrix{
(1+{1\over2}\psi{\bar\psi})\sigma_0&\psi\sigma_1\cr
{\bar\psi}\sigma_1&(1-{1\over2}\psi{\bar\psi})\sigma_0\cr
}
\label{symmetry}
\end{eqnarray}
with Grassmann variables $\psi$ and ${\bar\psi}$. It is important to notice
that the Dirac mass does not break this symmetry but only the term 
proportional to $\epsilon$ in (\ref{ssa0}), since $U^2$ is not the unit matrix.
Therefore, the field conjugate to the symmetry breaking field, which is 
quadratic in the superfield, must be studied in order to take the relevant
symmetry properties into account.
This leads to the collective field representation \cite{zie6,efetov}
\begin{equation}
\pmatrix{\chi_r{\bar\chi}_r&\chi_r{\bar\Psi}_r\cr
\Psi_r{\bar\chi}_r&\Psi_r{\bar\Psi}_r\cr
}\leftrightarrow
{\bf Q}_r=\pmatrix{
Q_r&{\bar\Theta}_r\cr
\Theta_r&-iP_r\cr
}.
\end{equation}
The matrix elements $Q_r$,...,$P_r$ are $2\times2$ 
%Hermitean
matrices, corresponding to the two components of $\Psi_r$ and $\chi_r$.

Since the Dirac Hamiltonian
$H_D$ is usually obtained from a large scale (or low energy) approximation of
a non-relativistic problem, there are several ways to introduce disorder which
are motivated by the original condensed matter systems. One starting point is,
for instance, the network model of Chalker and Coddington
\cite{chalk} for the QHT. This phenomenological description implies a random
Dirac mass, a random energy term and a random vector potential
\cite{ho/chalker}.
Thus this model represents a complex situation which also includes fluctuations
of the magnetic field. Here we are interested only in the simplest possible
case for the QHT of a system in a homogeneous magnetic
field. (Strong fluctuations of the random vector potential may drive the
system into
another universality class. At this point it is not clear if randomness in
the vector potential is relevant in the experiments on a 2D electron gas.)
The QHT can also be described by a tight-binding
model with a homogeneous magnetic field \cite{ludwig} in a random
chemical potential. The latter would lead to a random Dirac mass. However,
there was the argument that the random Dirac mass alone does not present
the generic situation for the QHT because the DOS is zero
at low energy \cite{ludwig}, i.e. there are no bulk states even in the
presence of disorder. It turned out though that these states exist if
one goes beyond perturbation theory. This effect was also found in
numerical calculations \cite{HatsugaiLee,LeeWang}.
%, and the DOS is $\propto\exp(-\pi/g)$ \cite{zie2}.
A consistent
treatment of this nonperturbative contribution can be based on an effective
field theory derived from the collective field ${\bf Q}$ \cite{zie6}. This
representation will be used in the following to discuss the breaking of the
supersymmetry defined in (\ref{symmetry}) and its consequences for the
existence of delocalized states.
\\  

Averaging over a Gaussian random Dirac mass $M$ (where $\langle M_r\rangle =m$
and
$\langle M_rM_{r'}\rangle =g\delta_{rr'}$) and
transforming the functional integral to the collective field
creates the new action \cite{zie6,efetov}
\begin{equation}
S' = 
{1\over g}\sum_r{\rm Trg}_4({\bf Q}_r^2)+\ln{\rm detg}({\bf H}_0+i\epsilon
-2\tau{\bf Q}\tau)
\label{action1}
\end{equation}
with ${\bf H}_0=\langle {\bf H}_D\rangle$ and the $4\times4$ diagonal matrix 
$\tau=((\sigma_3)^{1/2},(\sigma_3)^{1/2})$. ${\rm Trg}_4$ and ${\rm detg}$
are the ``supertrace' and the `superdeterminant', respectively \cite{efetov}.
In particular, the two-particle Green's function at $r\ne r'$ then reads 
\begin{equation}
K(r,r';\epsilon)=g^{-2}\langle(\Theta_{r,12}+\Theta_{r,21})
({\bar\Theta}_{r',12}+{\bar\Theta}_{r',21})\rangle_Q.
\end{equation}
The functional integral $\langle...\rangle_Q=\int...\exp(-S'){\cal D}{\bf Q}$
can be approximated by a saddle point integration.
%%%%%%%%%%% Saddle Point Expansion %%%%%%%%%%%%%%%%%%%%%%%%%%%%%%%%%%%%%%
A special saddle point is ${\bf Q}_0=(m/4)\gamma_0-i(\eta/2)\gamma_3$, where
$\gamma_j$ is the diagonal block matrix $(\sigma_j,\sigma_j)$
and $\eta=\pi g\rho$ is proportional to the average DOS $\rho$ \cite{zie6}. 
The symmetry transformations are now applied to the saddle point solution 
$\tau{\bf Q}_0\tau$.
The discrete transformation (\ref{discrete}) changes the sign of $\eta$,
i.e. the discrete symmetry of the massive Dirac Hamiltonian is spontaneously
broken if $\lim_{\epsilon\to0}\eta\ne 0$.
This is the case for $-m_c<m<m_c$ with $m_c=2\exp(-\pi/g)$
\cite{zie6}.
The supersymmetry transformation (\ref{symmetry}), on the
other hand, gives $\eta\to \eta U^2$.
Thus $\lim_{\epsilon\to0}\eta\ne0$ indicates also a spontaneously
broken supersymmetry.
This behavior is analogous to a Heisenberg ferromagnet, where $\eta$ 
corresponds to the magnetization and $\epsilon$ plays the role of the 
external magnetic field. However, the situation is more complex for the Dirac
fermions because the breaking of two symmetries is involved, a supersymmetric
and a discrete one.
As a consequence of the supersymmetry of $S_0$ at $\epsilon=0$
there is not just an isolated saddle point but a whole saddle point manifold,
created by the symmetry transformation $U$. Therefore, the field
\begin{eqnarray}
{\bf Q'}_r & = & {\tilde U}_r{\bf Q}_0{\tilde U}^{-1}_r
%\tau^*U_r(-i{\eta\over2}\gamma_0+{m\over4}\gamma_3)U_r\tau^*
={m\over4}\gamma_0-i{\eta\over2}\tau^* U_r^2\tau^*
\nonumber\\
& = & {\bf Q}_0 -i\eta\pmatrix{
\psi_r{\bar\psi}_r\sigma_3&-i\psi_r\sigma_1\cr
-i{\bar\psi}_r\sigma_1&-\psi_r{\bar\psi}_r\sigma_3\cr
}
\label{manifold1}
\end{eqnarray}
controls the fluctuations around the saddle point manifold with
${\tilde U}_r=\tau^*U_r\tau$ and ${\tilde U}^{-1}_r=\tau U_r\tau^*$.
$U_r$ is here the matrix $U$ of Eq. (\ref{symmetry}) in which the Grassmann
variable $\psi$ is replaced by the Grassmann field $\psi_r$.
That means for the large scale properties the integration with respect to
${\bf Q}_r$ can be restricted to an
integration with respect to the field ${\bf Q'}_r$.
Thus the critical (long-range) part of the random mass Dirac theory
is controlled by the one-component fermion (Grassmann) field $\psi_r$.
The bosonic (complex) field has only short-range correlations and,
therefore, is localized by the disorder. The reason is that the
bosonic field corresponds to the {\it discrete} symmetry transformation
(\ref{discrete}) which has a long-range mode only at the critical point where
the order parameter $\eta$  vanishes. The latter is indeed the case because the
localization length of
$Q_{11}-Q_{22}$ and $P_{11}-P_{22}$ increases like $(m_c-|m|)^{-1/2}$
as the critical value $\pm m_c$ is approached from $|m|<m_c$ \cite{zie6}.
This indicates a growing influence of these bosonic fields on the large scale
properties. 

The expansion of (\ref{action1})
up to second order in the gradients yields in general an action of the type
\cite{wegner,efetov,altshuler,pruisken}
\begin{eqnarray}
{i\epsilon
}\int d^2r{\rm Trg}_4(\gamma_3{\bf Q'}_r)
+\alpha\int d^2r{\rm Trg}_4(\nabla{\bf Q'}_r\cdot\nabla{\bf Q'}_r)
\nonumber\\
-\beta\int d^2r\sum_{\mu,\nu}\epsilon_{\mu\nu}
{\rm Trg}_4({\bf Q'}_r\nabla_\mu{\bf Q'}_r\nabla_\nu{\bf Q'}_r),
\end{eqnarray}
where $\epsilon_{\mu\nu}$ is the antisymmetric unit tensor, and the 
parameters $\alpha$ and $\beta$ are determined by the model. In particular,
for the quantum Hall effect there is
$\alpha=\sigma_{xx}$, the (unrenormalized) longitudinal conductivity, and
$\beta=\sigma_{xy}$, the (unrenormalized) Hall conductivity \cite{pruisken}. 
The topological term 
$\int d^2r\sum_{\mu,\nu}\epsilon_{\mu\nu}{\rm Trg}_4({\bf Q'}_r
\nabla_\mu{\bf Q'}_r\nabla_\nu{\bf Q'}_r)$ 
takes care of the Hall plateaux because the latter are a consequence of
the (topological) edges states in the presence of {\it localized} bulk states.
At the QHT, however, transport is dominated by {\it delocalized} bulk states.
Therefore, the topological term should not play a crucial role in this case.
In fact, for the Dirac Hamiltonian $H_D$ with $m=0$,
i.e. for the choice ${\bf Q}'$ of Eq.(\ref{manifold1}), the topological
term vanishes. The only terms which remain in the action are the linear 
off-diagonal elements of ${\bf Q}'$
\begin{equation}
S''= 
%2\int ({\bf I}_k')_{22}{\bar \psi}_{k}\psi_{-k}d^2k
(1/\pi\rho)\int d^2r{\bar \psi}_r(\epsilon+D \nabla^2) \psi_{r},
\label{freeferm}
\end{equation}
where the average DOS $\rho$ and the diffusion coefficent $D$ can be
evaluated form the saddle point equation. 
This surprisingly simple result, which satisfies the Ward identity 
$K(q=0,\epsilon)=\pi\rho/\epsilon$, reflects the fact that only a one-component
Grassmann field contributes to the massless fluctuations, created by the broken
supersymmetry. That means there is a simple physical structure for
the well-delocalized Dirac fermions in the vicinity of $m=0$.
The divergent localization length of two real boson components
will eventually turn into a restoration of the supersymmetry, where 
$\lim_{\epsilon\to0}\eta=0$.
Since the supersymmetric theory in 2D does not
have delocalized states \cite{efetov}, the restoration
of the supersymmetry must be accompanied by a transition into a
localized regime. This is the regime characterized by the Hall plateaux.
The critical behavior of the two real fields (which can be considered as the
two components of one complex boson field) at $m=\pm m_c$ due to the
spontaneously broken discrete symmetry (\ref{discrete})
invalidates $S''$ near these points.
It must be replaced by a more complicated field theory which
includes both the critical boson field and the critical Grassmann
field of (\ref{freeferm}). This would require an additional matrix
field
\begin{equation}
\pmatrix{
q_r\sigma_3 & 0\cr
0 & -i p_r\sigma_3 \cr
},
\end{equation}
added to ${\bf Q'}_r$ in (\ref{manifold1}).
The real field components $q_r$ and $p_r$ are related to $Q_{11,r}-Q_{22,r}$
and $P_{11,r}-P_{22,r}$, respectively.

As a direct consequence of these results the value of the conductivity at 
$m=0$ (the ``conduction peak'') can be evaluated from the Einstein 
relation $\sigma_{xx}=(e^2/\hbar)D\rho$ \cite{zie/jug}
\begin{equation}
\sigma_{xx}={e^2\over\pi h}{1\over 1+g/2\pi}.
\label{cond}
\end{equation}
This is in agreement with experimental results \cite{mceuen,rokhinson}
and other theoretical work \cite{ruzin,cooper}.
For weak disorder the second factor can be neglected. In this case the peak
value is just the universal constant $\sigma_{xx}=e^2/\pi h$. The latter
was obtained for Dirac fermions in a random vector potential \cite{ludwig}
and for the lowest Landau level with random spin scattering \cite{hikami},
regardless of the strength $g$.
Thus the extra factor in Eq.(\ref{cond}) indicates that the 
random Dirac mass, representing a random scalar potential, has
a stronger effect on the Dirac particle than the random vector potential
or a random spin scattering.\\

In conclusion, the large scale behavior of the two-particle Green's function
of 2D Dirac fermions with random mass is characterized by a single massless
Grassmann field on the interval $[-m_c,m_c]$ of the average Dirac $m$.
It describes delocalized states with non-singular DOS due to a broken 
supersymmetry
by a diffusion propagator. There is a two-component real bosonic field which
has a divergent localization length as the critical points $\pm m_c$ are
approached. It corresponds to a broken discrete symmetry of the Dirac
Hamiltonian.
This mechanism of delocalization is different from the one which is
responsable for delocalized states in the random vector potential or
for random spin scattering.


\begin{references}
\bibitem{note}
It should be noticed, however, that the standard version of the quadratic
form in Eq. (\ref{ssa0}) can be obtained by renaming the Grassmann
field as $\Psi_r\to i{\bar\Psi}_r$, ${\bar\Psi}_r\to i\Psi_r$.
\end{references}
\end{document}